\documentclass[journal]{IEEEtran}
\usepackage{multirow}
\usepackage{amsfonts}
\usepackage{amsmath}
\usepackage{color}
\usepackage{bm}
\usepackage{amssymb}
\usepackage{xcolor}
\usepackage{graphicx}
\usepackage[ruled,vlined,linesnumbered]{algorithm2e}
\usepackage{fancyhdr}
\usepackage{epstopdf}
\usepackage{paralist}
\usepackage{url}
\usepackage{subfigure}
\usepackage{lipsum}
\usepackage{xcolor,colortbl}
\usepackage{comment}
\usepackage[official]{eurosym}
\usepackage{cite}
\usepackage{booktabs}
\usepackage{subfig}
\IEEEoverridecommandlockouts

\makeindex

\begin{document}

\title{Supervised Learning based QoE Prediction of Video Streaming in Future Networks: A Tutorial with Comparative Study}
 \author{\IEEEauthorblockN{Arslan Ahmad, Atif Bin Mansoor, Alcardo Alex Barakabitze, Andrew Hines, Luigi Atzori and Ray Walshe}

\thanks{Arslan Ahmad is with Cardiff School of Technologies, Cardiff Metropolitan University, United Kingdom, Email: aahmad@cardiffmet.ac.uk}
\thanks{Atif Bin Mansoor is with University of Western Australia, Australia. Email: atif.mansoor@uwa.edu.au}
\thanks{Alcardo Alex Barakabitze and Ray Walshe are with Dublin City University, Ireland. Email: [alcardo.barakabitze, ray.walshe]@dcu.ie} 
\thanks{Andrew Hines is with  University College Dublin, Ireland. Email: andrew.hines@ucd.ie}
\thanks{Luigi Atzori is with  University of Cagliari, Italy. Email: l.atzori@ieee.org} 
  }
\maketitle

\begin{abstract} 
The Quality of Experience (QoE) based service management remains key for successful provisioning of multimedia services in next-generation networks such as 5G/6G, which requires proper tools for quality monitoring, prediction and resource management where machine learning (ML) can play a crucial role. In this paper, we provide a tutorial on the development and deployment of the QoE measurement and prediction solutions for video streaming services based on supervised learning ML models. Firstly, we provide a detailed pipeline for developing and deploying supervised learning-based video streaming QoE prediction models which covers several stages including data collection, feature engineering, model optimization and training, testing and prediction and evaluation. Secondly, we discuss the deployment of the ML model for the QoE prediction/measurement in the next generation networks (5G/6G) using network enabling technologies such as Software-Defined Networking (SDN), Network Function Virtualization (NFV) and Mobile Edge Computing (MEC) by proposing reference architecture. Thirdly, we present a comparative study of the state-of-the-art supervised learning ML models for QoE prediction of video streaming applications based on multiple performance metrics. 
\end{abstract}
\begin{IEEEkeywords}
Quality of Experience (QoE), QoE prediction, QoE measurements, Machine Learning, Video streaming, Software-Defined Networking (SDN), Network Function Virtualization (NFV).
\end{IEEEkeywords}
\section{Introduction}
\label{sec:intro}

The Internet traffic is predominantly multimedia traffic due to the highly increasing demand for Over The Top (OTT) services such as Facebook, YouTube, Skype, etc. According to the Cisco Visual Networking Index (VNI)~\cite{Cisco2017}, the IP video traffic will represent 82\% of all IP traffic by 2022. 
Due to the ever-increasing demand for multimedia services, especially video streaming, Internet Service Providers (ISPs)/Mobile Network Operators (MNOs) need to incorporate novel network traffic management and monitoring solutions~\cite{skorin2018survey}.

The Quality of Experience (QoE) is a multidimensional concept which measures the quality perceived by the end-user depending upon multiple influence factors such as application, network, business, system (user device) and context~\cite{barakabitze2019qoe}.
The QoE of the 
HTTP Adaptive Streaming (HAS)-based video streaming depends on multiple application and network Key Performance Indicators (KPIs)~\cite{ahmad2020timber}. The deployment of advanced QoE monitoring and management solutions for video streaming services can leverage network softwarization/virtualization technologies such as Software Defined Networks (SDN) and Network Function Virtualization (NFV) adopted in 5G networks and beyond. In this context, Machine Learning (ML) models can be deployed to provide QoE prediction, monitoring and measurement solution. 
Today, the OTT video streaming services are end-to-end encrypted making it difficult for the MNO/ISP to deploy QoE monitoring and prediction solutions (as no or few QoE related Key Quality Indicators (KQIs) information is available to MNO/ISP) for the next-generation networks management~\cite{ahmad2019mno,ahmad2017ott}.
The studies in~\cite{wassermann2020vicrypt,orsolic2020framework} propose the QoE prediction for the encrypted video streaming services but still, it is difficult to have an accurate prediction of the QoE from the encrypted video streaming traffic. However, this paper focuses on data-driven ML-based QoE prediction of video streaming with the perspective of network management in the 5G networks using OTT-MNO/ISP collaboration for information exchange.  Accordingly, in the considered framework, when QoE monitoring/prediction solution is deployed on the ISP/MNO side using SDN and NFV for QoE-aware network management, QoE related information is retrieved from the OTT client.   
In this paper, we investigate ML-based QoE prediction for video streaming applications where we provide a tutorial and comparative study. The contribution of this work is as follow: 
\begin{itemize}
    \item We present the step-by-step procedure to design, develop and deploy supervised learning-based video streaming QoE prediction algorithms.
    \item Secondly, we propose a reference architecture and discuss the deployment of the ML model based QoE prediction and measurement solution for QoE management of video streaming services using network enabling technologies such as SDN, NFV and MEC in the next generation networks 5G/6G.
    \item Thirdly, we present a comparative analysis of the seven state-of-the-art supervised learning ML models for QoE prediction of video streaming applications based on multiple performance metrics.
     \end{itemize} 

The paper is structured as follows: Section~\ref{sec:tutorial} provides procedural guidelines for developing ML-based QoE prediction model. Section \ref{sec:MLforQoEprediction} presents a reference architecture for QoE prediction/management. Section~\ref{sec:experimentssetup} provides the details of experiments for comparative study while Section~\ref{sec:comparative-analysis} provides a comparative analysis of the supervised learning algorithms using various performance metrics. Finally, Section~\ref{sec:con} concludes the paper.

\section{QoE Prediction using Supervised Learning}\label{sec:tutorial}
\begin{figure}[!t]
\centering
\includegraphics[width=\columnwidth]{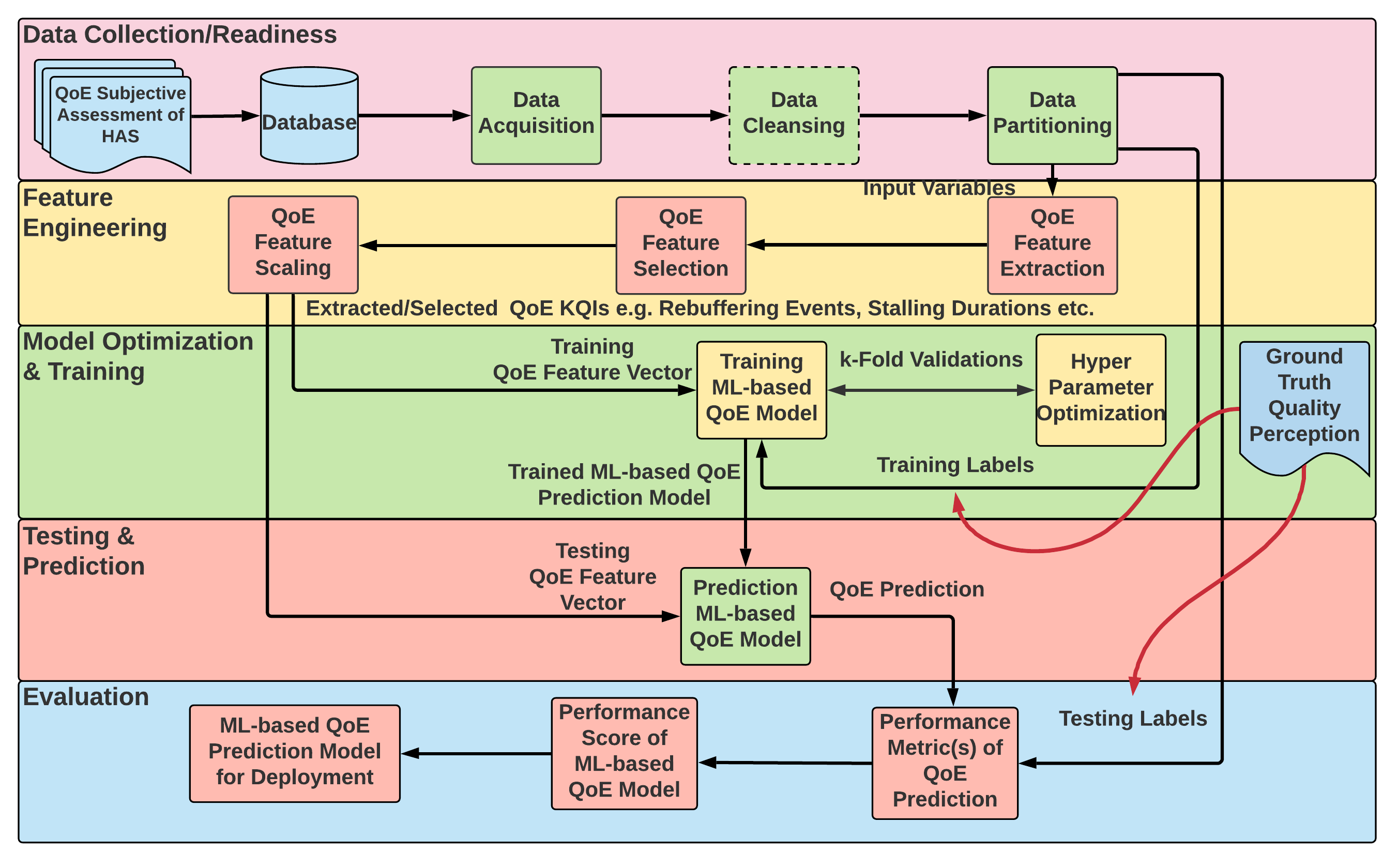}
\caption{Process flow diagram for QoE prediction of video streaming service using supervised learning.}
\label{fig:method}
\end{figure}

There are several stages of QoE prediction of video streaming services using ML which highly depends on the type of ML model~\cite{vega2018review}.
This section provides guidelines for developing supervised learning models for QoE prediction of HAS.
Fig.~\ref{fig:method} represents the sequence flow diagram for developing ML-based QoE prediction model of HAS. 

To acquire data, subjective assessments are conducted either in a controlled environment/lab or with crowdsourcing where influencing factors (input variables) of QoE are monitored together with the user-perceived quality (ground truth labels) forming a database of subjective quality perceptions~\cite{barman2019qoe}. 
In data collection/readiness stage, raw data comprising of input variables and ground truth are acquired and processed further to remove missing entries (optional) followed by the data partitioning to split data into training and testing subsets. 
The QoE features of HAS (number of rebuffering events, stalling durations etc.) are extracted/selected from the input variables for both training and testing phases in feature engineering stage. The extracted/selected QoE features are then scaled by the QoE feature scaling/normalization module. 
In model optimization \& training stage, the QoE feature vector of the training subset is fed to the ML model for the training ML-based QoE model. Every ML model has the hyper-parameters configuration which needs to be optimized according to the problem and the data.
In the testing and prediction stage, the trained ML-based QoE prediction model is provided with the only testing QoE feature vector for the QoE prediction. 
In the evaluation stage, the QoE predicted by ML model is evaluated in comparison with the testing labels based on QoE prediction performance metric(s). Once the trained ML-based QoE prediction model meets the performance criteria 
then it can be deployed for QoE-aware network management as highlighted in Section~\ref{sec:MLforQoEprediction}.

\begin{figure*}[!t]
\centering
\includegraphics[width=\textwidth]{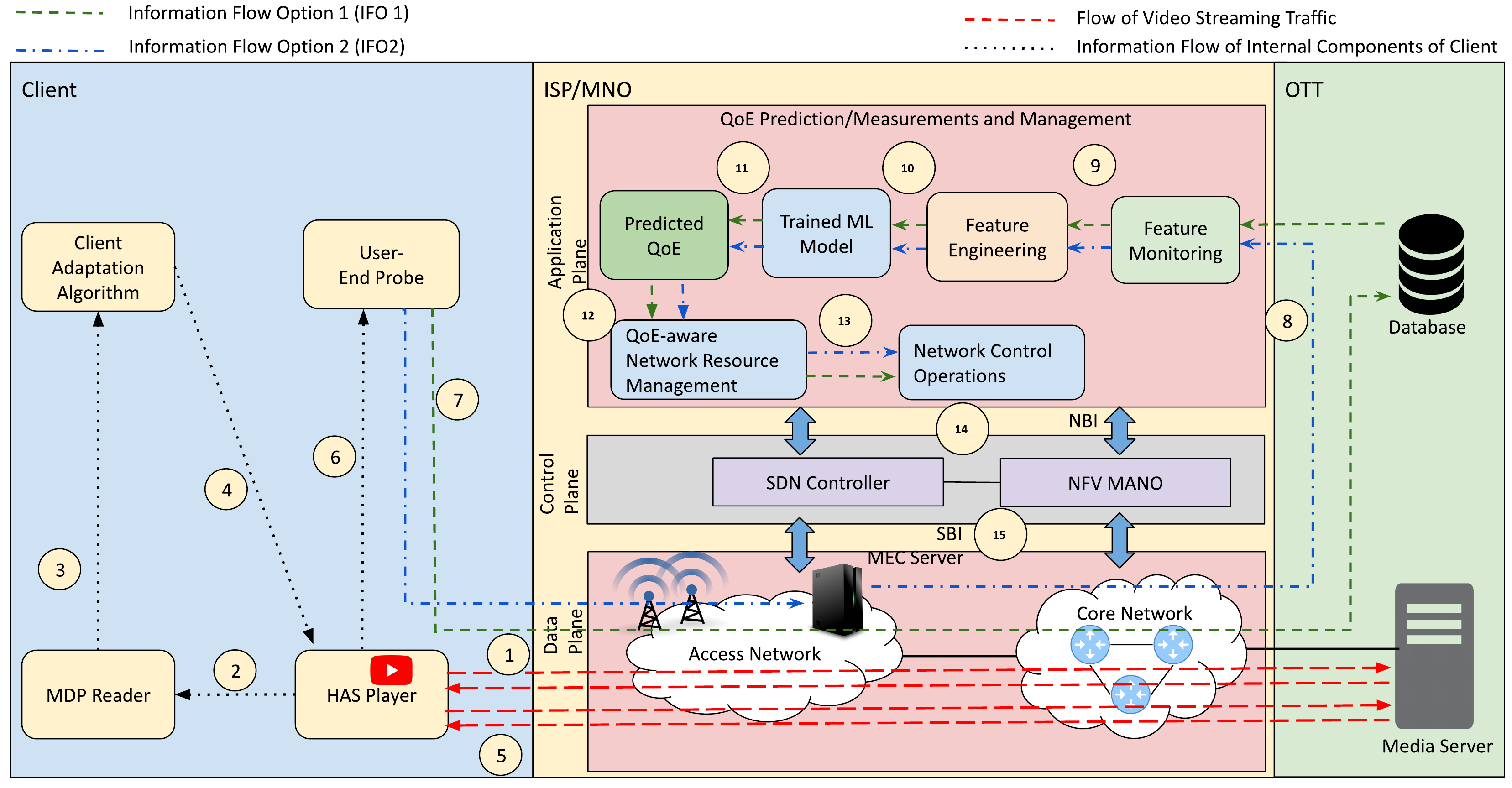}
\caption{Architectural components, implementation and deployment options using ML.}
\label{fig:MLAlgorith}
\end{figure*}

\section{Reference architecture for QoE Measurement and Management in Future Networks}\label{sec:MLforQoEprediction}
This section provides a reference architecture for QoE prediction/measurement and management in future networks using ML. The architecture shown in Fig.~\ref{fig:MLAlgorith} indicates three main actors in the end-to-end multimedia networking, namely the client, ISPs/MNOs and OTT services providers.
The reference architecture considers that OTT and ISP/MNO collaborate for QoE-aware network management where OTT collects QoE KQIs from the user-end probe based on users' preference/GDPR policy and shares it with ISP/MNO for ML-based QoE prediction and network resource optimization.
\subsection{OTT} 
In the reference architecture, the OTT offers a streaming media service directly to customers and businesses via the Internet. The media server contains multimedia content that can be delivered to client devices. The Database stores the QoE KQIs such as re-buffering event frequency, re-buffering event average duration, representation quality switching rate, video bitrates etc.
\subsection{Client}
The client-side includes components such as 
HAS Player, Client Adaptation Algorithm, 
Media Presentation Description (MPD) Reader and User-End Probe. The HAS player requests video content from the media server. In the HAS video streaming, the video has multiple representations in different coding bitrates and resolutions. To avoid stalling events during the video playback, the HAS client-side adaptation algorithm optimizes the client player bitrates based on the available network KPIs e.g. network throughput. During video streaming, the MPD reader provides information about the available video representations (e.g., number of video segments/chunks, video resolution, bitrates etc.). The user-end probe is responsible to collect the QoE KQIs and send them to the database/Multi-access Edge Computing (MEC) server. The HAS player implemented by OTT should allow ISP/MNO to collect/store this information to database/MEC server by user end probe using request–response protocols such as Remote Procedure Calls (RPC).
\subsection{ISP/MNO}
The reference architecture at the ISP/MNO side consists of three planes, namely the Data Plane, Control Plane and Application Plane.
\subsubsection{Data Plane}\label{dplne}
The Data Plane represents the ISP access network and core network. The data plane forwards packets from source to destination where SouthBound Interface (SBI) defines communication between control and data plane.
\subsubsection{Control Plane}\label{cnpln}
The control plane consists of an SDN controller and NFV Management and Orchestration (MANO) \cite{Barakabitze2020}. The NFV MANO is integrated into the reference architecture to offer greater flexibility for network functions deployment and dynamic network operation as well as QoE-based service provisioning. The SDN controller enables automation, softwarization and programmability of network resources, network configuration and QoE-optimization to improve network performance by the deployment of advanced applications such as QoE prediction/measurement and management.
\subsubsection{Application Plane}\label{mngplan}
The application plane implements six modules: the Feature Monitoring, Feature Engineering, Trained ML Model, Predicted QoE, QoE-aware Network Resource Management and Network Control Operations. The feature monitoring module is responsible for monitoring the QoE features/KQIs during a video streaming session periodically depending on monitoring frequency. The monitoring frequency/time-period highly depends upon the time-period for network resource optimization, accuracy/data requirements of optimization algorithm(s), the dataset used for training ML-based QoE prediction model and, the trade-off between the control plane traffic and monitoring frequency (higher monitoring frequency leads to higher control plane traffic). The study in~\cite{ahmad2019towards} shows that higher monitoring frequency for information exchange between OTT and ISP for QoE management of HAS services using SDN may result in higher delivered QoE and network reliability but monitoring frequency higher than 1/4 Hz may not improve delivered QoE further. Therefore, finding the optimal frequency for QoE/feature monitoring remains an open challenge. The feature engineering module extracts/selects the QoE feature vector from the monitored KQIs. The selected/extracted features are provided to the Trained ML Model which predicts the QoE based on the QoE feature vector. 

The NorthBound Interface (NBI) enables communications between the control plane and application plane.  
The output of the trained ML model, the estimated QoE is reported to the QoE prediction module which stores the predicted/measured QoE data and make it available to the QoE-aware Network Resource Management module for optimizing network resource based on which Network Control Operation module performs network-wide control actions.
\subsection{Information Flow and Deployment Options}
The reference architecture consists of two sequence flow options as shown in Fig.~\ref{fig:MLAlgorith}: Information Flow Option 1 (IFO1) indicates the flow of information via a
database on the OTT side to equip ISP/MANO with QoE KQIs. Information Flow Option 2 (IFO2) shows the flow of information through the MEC server which is integrated at the Radio Access Network (RAN). 
Information Flow of Internal Components of the Client remains common in both IFO1 and IFO2. The information flow sequence follows the following steps:
\begin{itemize}
 \item \textbf{Step 1: }The client requests video content from the media server using HTTP. The media server replies to the client with the requested MPD file.
 \item \textbf{Step 2: }The requested MPD file is parsed by the MPD reader.
\item \textbf{Step 3: }The MPD reader provides the client adaptation algorithm with the information about the requested video content characteristics (number of video segments, resolution etc.) in JSON format.
\item \textbf{Step 4: }The client adaptation algorithm adapts the bitrates for the requested video segments based on the current state of the network and provides feedback to HAS player during the video streaming session. 
\item \textbf{Step 5: }Based on the received feedback from the client adaptation algorithm, the HAS player keeps on requesting the video segments from the media server.
\item \textbf{Step 6--8: }The user-end probe collects the QoE KQIs from HAS player. The collected information (QoE KQI/input data) is then sent to the database on the OTT side in case of IFO1 or to MEC server in case of IFO2, that is utilized by the feature monitoring module.
\item \textbf{Step 9--11: }The QoE prediction/measurements module performs the following functionalities (a) the retrieval of the QoE KQIs by feature monitoring module, (b) application of feature engineering to the input data to get the relevant feature vector, (c) prediction of QoE via trained ML model using feature vector from feature engineering module and, (d) storing predicted QoE to be used by the other network management applications.
\item \textbf{Step 12--13:} The QoE-aware network resource management module performs optimization and management of the available network resources by utilizing the predicted QoE as an objective function based on network management policies and Service Level Agreements (SLAs).
\item \textbf{Step 14--15: }The network control operations module performs network operations such as network slicing in softwarized/virtualized networks, traffic flows engineering and network services configuration for QoE-aware allocation of the network resources. 

\end{itemize}
\subsection{QoE-aware Network and Service Management}

The proposed reference architecture can provide QoE-aware network resource management by exploiting predicted QoE to enable network management operations such as end-to-end network slicing, scaling of Virtual Network Functions (VNF), network traffic management, service prioritization and radio resource allocation using SDN and NFV. 
To exploit predicted QoE for network management, the proposed reference architecture can be used in following use cases: 
\begin{itemize}
    \item \textbf{Use Case 1: QoE-aware Network Slicing/Flow priority/traffic engineering --} The predicted QoE using ML models deployed on top of SDN controller can be utilized for performing and automating QoE-aware Multiprotocol Label Switching (MPLS) based flow prioritization, traffic re-routing and network slicing operations where predicted QoE can be translated to network QoS using QoE to QoS models~\cite{barakabitze2019qoe,Barakabitze2020}. For example, the works in~\cite{ahmad2019towards,ahmad2020timber} proposed using the QoE prediction model for driving QoE-aware network slicing/MPLS operations by considering predicted QoE as an objective function at the application layer of SDN controller which performs the network management operations at data plane using OpenFlow SBI. Moreover, the QoE prediction model deployment on top of NFV MANO may lead toward QoE-aware network automation for scaling/lifecycle management of VNF computational resources in the virtualized network infrastructure while saving CAPital EXpenditures (CAPEX)/OPerating EXpenses (OPEX) for the network providers. Furthermore, the predicted QoE can also drive the content distribution/management across MEC based surrogate servers in the ISP network domain to lower end-to-end content retrieval latency as shown in~\cite{ahmad2017ott}.   
    \item  \textbf{Use Case 2: QoE-aware radio resource allocation --} The QoE-aware optimization of the radio resources in the 5G networks requires the quality prediction model integration into the radio network management. For example, in~\cite{ahmad2019mno}, the study proposed a zero-rated QoE approach where OTT and MNO collaborate to share QoE KQIs where Physical Resource Blocks (PRBs) at RAN are allocated to users based on predicted QoE and relating the video KQIs with network QoS KPIs. Therefore, at the RAN, the predicted QoE may allow effective radio resource allocation at the scheduler within Medium Access Control (MAC) layer.
    With the recent standardization advances towards softwarization and virtualization of RAN at 3GPP, O-RAN and ETSI,  the AI/ML-driven QoE prediction model can be deployed at the application layer of software-defined RAN (SD-RAN) controller as well as ETSI MANO. For example, the RAN softwarization efforts at O-RAN W2 and W3 consider the deployment of QoE optimization through SD-RAN controller at near real-time and non-real-time optimization interval. The deployed QoE prediction model on top of the SD-RAN controller in the next-generation network can allow QoE-based optimization of the link capacity, mobility management and admission control at RAN by switching/combining multiple access nodes/underlying technologies (e.g. 5G New Radio (NR), LTE, WiFi) using multi-Radio Access Technologies (multi-RAT).  Furthermore, the predicted QoE can also be utilized as an objective function for VNF placement and radio protocol layers split optimization during virtualized RAN lifecycle management using ETSI MANO architecture.   
\end{itemize}

\section{Experiments Settings for Comparative study}\label{sec:experimentssetup}
This section discusses the settings of the experiments for the comparative study.
All the experiments are performed on a Python-based ML development environment using \textit{Scikit-Learn, Pandas and Seaborn} libraries installed on Ubuntu 16.04 workstation with $16$ GB RAM and corei7 CPU. All the results are 5-fold cross-validated. 

\subsection{Data Collection/Readiness}
The experiments utilize the largest publicly available database with QoE based ground truth provided by Duanmu et al.~\cite{duanmu2018sqoe}. 
This database has $450$ sample size with the key QoE influencing factors of HAS video streaming with ground truth labels.  
Each sample is represented by a video sequence of average duration of $13$ seconds with the subjective score (ground truth), objective QoE metrics and QoE KQIs. The video streaming sessions within the database are generated by six different client-side adaptive algorithms under 13 different network bandwidth conditions which are evaluated by 34 subjects.
For the experiments, the data doesn't contain any missing values so data cleansing was not applicable. The data samples are randomly divided into training and testing data subsets with $80/20$ ratio, respectively.   

\subsection{Feature Engineering}
For the experiments, feature scaling is performed by standardizing the extracted features to the normal distribution such that each feature has mean ($\mu=0$) and standard deviation ($\sigma=1$). 
Some of the features used in our experiments are inspired by the QoE parameter used in ITU-T P.1203 standard~\cite{P1203}, which is a parametric bit-stream based quality assessment/prediction model for HAS. The ITU-T P.1203 has four different modes of operation where different QoE KQIs for HAS are considered such as the total number of stalling events, stalling duration, video coding quality etc. 
We consider the following key features which are extracted/selected from the input variables:
\begin{enumerate}
    \item \textit{Initial video loading time} is the initial loading time taken by the video streaming service to load video segments at the client buffer before starting the playback.
    \item \textit{Total number of stalling events (except initial video loading)}, which represent the player buffer starvation conditions during the HAS video streaming which cause video playback to be interrupted. The total number of stalling events is a crucial QoE influence factor~\cite{P1203}. 
    \item \textit{Total stalling duration} is the sum of the duration of the stalling events in seconds that occurs during the video playback~\cite{P1203}.
    \item \textit{Stalling frequency} is a measure of how often stalling events happen during the video playback which is also considered in ITU-T P.1203 standard~\cite{P1203}. 
    \item \textit{Stalling ratio} is also an important QoE influence factor according to ITU-T P.1203~\cite{P1203}. It represents the quality degradation effect on the user perception based on the total duration of stalling events over the length of the video playback. 
    \item \textit{Time of the last stalling from playback end} incorporates the recency effect of the human perception related to QoE degradation~\cite{P1203}. It is computed by the time stamp difference between the video playback length from the occurrence time of the last stalling event.
    \item \textit{Playback bitrates} represent the visual quality of the video representation. The average playback bitrate is considered highly correlated to the user-perceived QoE~\cite{duanmu2018sqoe}.  
    \item \textit{Video encoding frame rate} is considered to take into account video encoding setting in the features.  
    \item \textit{Video quality layer} is considered due to the adaptive nature of the HAS video streaming. We consider the median of the video quality layers being played during video playback. 
    \item \textit{Visual Quality Index}
    is proposed in the ITU-T P.1203 standard to compute the visual quality index over the video session on the ITU-T ACR scale ($1-5$) as a function of the playback bitrates, device resolution, and video encoding resolutions~\cite{P1203}. 
\end{enumerate}
\subsection{ML Model Optimization \& Training}
The following seven well known supervised ML models are considered: Support Vector Regression (SVR), Random Forest (RF), Gradient tree Boosting (GB), Stochastic Gradient Descent (SGD), Multi-Layer Perceptron (MLP) based Neural Network (NN), K-Nearest Neighbour (K-NN) and Decision Tree (DT).
During the training stage, hyper-parameter optimization with 5-fold cross-validation is performed using the grid search algorithm using $R^2$ as an objective criterion.
In addition to the default parameters of the above-mentioned ML models in \textit{Scikit-Learn} documentation~\cite{scikit-learn}, the following parameters are selected as a result of hyper-parameter optimization: 1) SVR has RBF kernel with the coefficient of regularization (C) equal to $10$; 2) RF has $0$ random state and $500$ estimators; 3) GB has $0.01$ learning rate and  $500$ estimators; 4) SGD has $1000$ maximum epochs with $1e-3$ tolerance limit; 5) NN has $0.001$ learning rate with the rectified linear unit as activation function and $20$ hidden layers; 6) K-NN has $10$ number of neighbours with uniform weights and 7) DT has $0$ random state with $9$ minimum number of samples required to split an internal node.

\subsection{ML Model Evaluation and Performance Metrics}
The evaluation stage considers five different performance metrics for ML models: 
Mean Square Error (MSE), Root Mean Square Error (RMSE), Mean Absolute Error (MAE), Coefficient of Determination $R^2$, Pearson Linear Correlation Coefficient (PLCC), Spearman's Rank Correlation Coefficient (SRCC).
The PLCC and SRCC correlate the predicted QoE with actual user-perceived quality in the ground truth data i.e., high PLCC and SRCC are desirable.  
\section{Comparative Analysis}\label{sec:comparative-analysis}
This section provides a comparative analysis of the supervised learning based QoE prediction models.

\begin{figure}[!t]
\centering
\includegraphics[width=0.9\columnwidth]{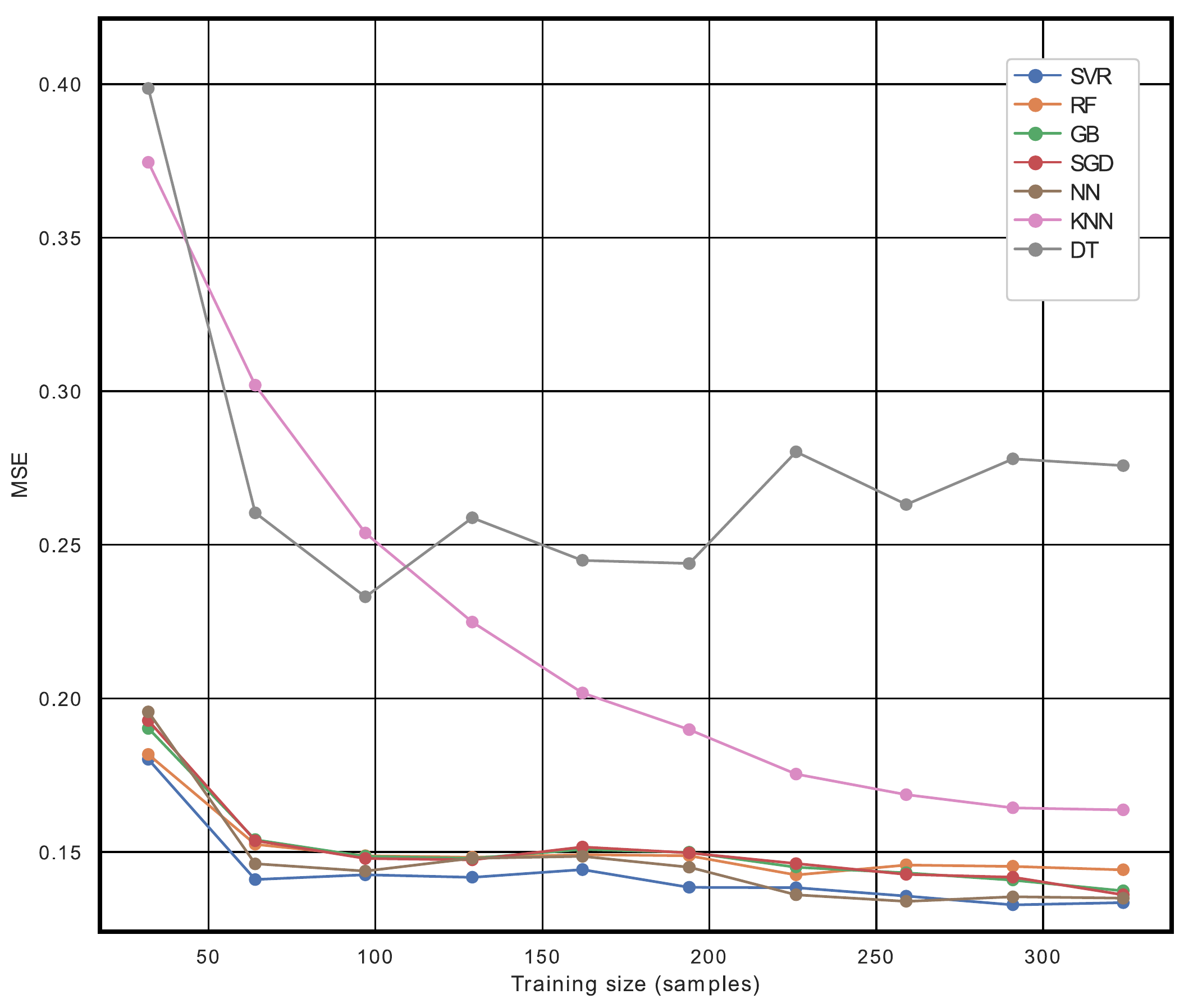}
\caption{Learning curve of ML models: $MSE$ vs training size.}
\label{fig1}
\end{figure}
\subsection{Learning Curve}\label{sec:learning-curve}
Learning curve comments on the evolution and convergence of the supervised learning ML model based on minimization of the objective function (here we consider 
$MSE$ as an objective function). 
Fig.~\ref{fig1} provides a comparison of ML algorithms based on the $MSE$ by varying training size.
The training size is varied with the step size of $10\%$ of the training data. The SVR, RF, GB, NN and SGD ML models show the minimum $MSE$ even for the small training data set size, which keeps on decreasing with the increase in the training samples size. While the K-NN and DT show higher $MSE$ for the small training size as compared to other ML models, which decreases exponentially with the increase in the training size. 
Therefore, based on the learning curve for $MSE$, SVR, RF, GB, NN and SGD show the best performance as compared to K-NN and DT. 
%
Thus, for the small training data size, ML models such as SVR, RF, GB, NN and SGD will be a good choice for QoE prediction of HAS video streaming.  
\begin{figure}[!t]
\centering
\includegraphics[width=0.85\columnwidth]{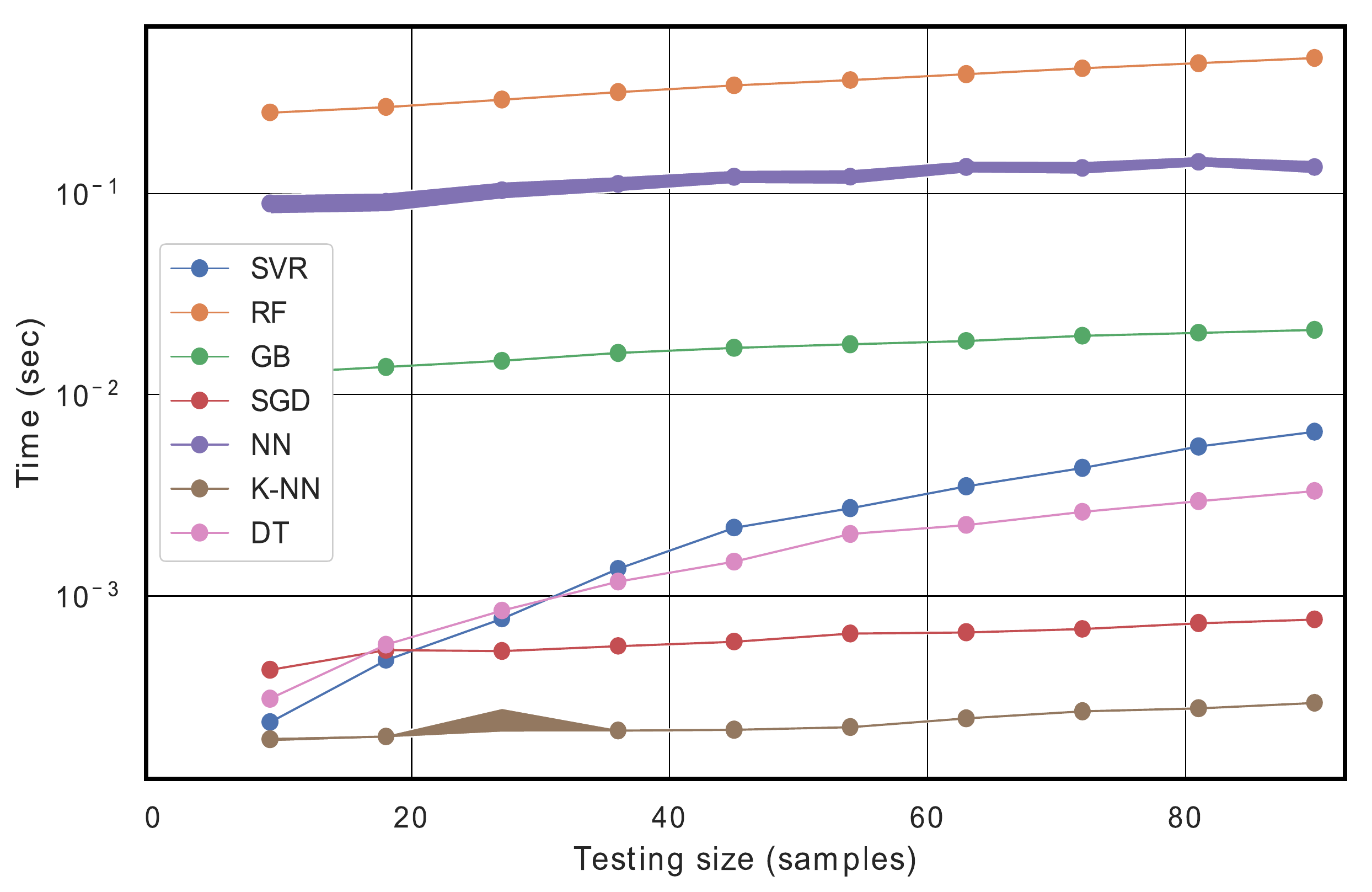}
\caption{Execution time of the ML algorithms: Testing time vs testing size.}
\label{fig:time}
\end{figure}
\subsection{Execution Time}\label{sec:execution-time}
The execution time remains important for real-time applications especially when it comes to real-time QoE prediction as the solution is required to be scalable as well as computationally less expensive.
Fig.~\ref{fig:time} provides a comparison of the ML models based on the execution time for the testing phase once the model is already trained. For the computation of the execution time, $100$ runs of the experiments are performed and the mean execution time of the ML models is reported in Fig.~\ref{fig:time}. In this case, RF takes the highest execution time while K-NN shows the lowest execution time. 
The execution time for NN and RF varies slightly with the increment in data size, whereas DT and SVR execution time increase with the increase in testing data size. The RF model is computationally more expensive as compared to the other ML models based on execution time. However, the RF model is still scalable as execution time slightly changes with the increment in data size. Furthermore, the testing execution time of SVR and DT highly depends on the size of the data i.e., the SVR and DT become more computationally expensive with the increase in the data size.


\subsection{Prediction Metrics}\label{sec:prediction-metrics}

Table~\ref{tab:summary} provides a summary of comparative analysis of the machine learning approaches based on the performance metrics. In terms of RMSE and MAE, RF predicts the QoE with the minimum RMSE and MAE while DT shows the highest RMSE and MAE. We noticed that for the QoE prediction, the complex ML models based on Ensemble Methods (where several ML models are combined to form optimal and complex ML model e.g. RF is made by bagging multiple DT together) such as RF and GB perform better as compared to simple ML models such as DT or K-NN. The major reason is that QoE prediction is a complex non-linear problem and complex ML models based on Ensemble Methods outperforms other models. Similarly, in the case of $R^2$, RF and GB represent the highest score as compared to other algorithms.  
Among all considered ML algorithms, the QoE predicted by the RF algorithm has the highest PLCC and SRCC while DT shows the lowest scores for PLCC and SRCC. The QoE prediction by GB, SVR, SGD and NN ML models also depict high PLCC and SRCC scores.




\begin{table}[t]
\caption{Comparison of the supervised learning approaches.}
\label{tab:summary}
\resizebox{\columnwidth}{!}{%
\begin{tabular}{l|lllllll}
\toprule \toprule
\textbf{Metrics}                & \textbf{SVR}   & \textbf{RF} & \textbf{DT} & \textbf{GB} & \textbf{K-NN}  & \textbf{NN} & \textbf{SGD} \\ \toprule \toprule
RMSE & 0.125          & \textbf{0.106}         & 0.166         & 0.118                   & 0.145          & 0.133              & 0.120        \\ \midrule
MAE   & 0.270          & \textbf{0.262}         & 0.319                  & 0.274                   & 0.301 & 0.276              & 0.272        \\ \midrule
$R^2$                    & 0.773          & \textbf{0.806}         & 0.698         & 0.786                   & 0.736          & 0.758            & 0.781        \\ \midrule
PLCC                  & 0.882          & \textbf{0.902}         & 0.853         & 0.890                   & 0.861          & 0.880        & 0.887        \\ \midrule
SRCC                  & 0.876          & \textbf{0.898}         & 0.859         & 0.884                   & 0.856          & 0.883        & 0.881        \\ \bottomrule \bottomrule
\end{tabular}}
\end{table}

\section{Conclusion}
\label{sec:con} 
In this paper, we provide a tutorial and comparative study on supervised learning ML models based on QoE prediction in the future networks. In the tutorial part, we discussed the deployment of ML models for the QoE prediction of multimedia streaming using network enabling technologies such as SDN, NFV, MANO and MEC in the next generation networks by providing a reference architecture. We provided a detailed tutorial on supervised learning ML model pipeline for QoE prediction of video streaming services. For the comparative study, experiments are performed where the key QoE influencing factors for video streaming are considered. The comparative analysis has provided performance comparison of the ML models based on learning curves, training/testing execution time, $R^2$, root mean square error, median absolute error, mean absolute error, PLCC and SRCC. The experiments conducted for the comparative study depend on the dataset of short duration video sequences (13-second average duration) which may limit the comparative analysis towards only short video sequences. Thus, future works are required for the construction of long video sequence dataset to development and comparison of ML-based QoE prediction models. However, this may not limit the applicability of the proposed reference architecture for the deployment of ML-based QoE prediction/management.

QoE prediction for video streaming in future networks will rely on careful collection and consideration of the data at every stage of the process illustrated in Fig.~\ref{fig:method}. We have highlighted these considerations and illustrated through experiments that even when supervised, data-driven models are provided with the same data and features, the robustness and accuracy can vary significantly. While many algorithms can potentially provide similar accuracy, other considerations such as training and execution times will influence the adoption of models in real networks. 
\section*{Acknowledgment}
This work is funded under SIP and GA fellowship by Cardiff Metropolitan University, United Kingdom, EU H2020 MSCA grant agreement No.
801522 and ADAPT Centre grant number 13/RC/2106\_P2.

\bibliographystyle{IEEEtran}
\bibliography{references}


\section*{Biographies}
ARSLAN AHMAD 
is currently working as Senior Lecturer at Cardiff Metropolitan University, United Kingdom.

ATIF BIN MANSOOR is currently working at the University of Western Australia, Austalia.

ALCARDO ALEX BARAKABITZE is currently a MSCA ELITE-S Fellow in the School of Computing, Dublin City University, Ireland. 

ANDREW HINES is a senior member of IEEE and an Assistant Professor with the School of Computer Science, University College Dublin, Ireland, where he leads the QxLab research group.

LUIGI ATZORI is senior member of IEEE and Professor at the Department of Electrical and Electronic Engineering at the University of Cagliari (Italy), where he leads the laboratory of Multimedia and Communications (MCLab). 

RAY WALSHE is a Senior IEEE member and Assistant Professor in the School of Computing, Dublin City University (DCU), Ireland with 20 years of industrial/academic experience.

\end{document}